\documentstyle[aps,preprint,epsf]{revtex}
\tighten
\begin{document}
\draft

\title  {
Boson mappings and four-particle correlations in algebraic
neutron-proton pairing models
}

\author  {J. Dobe\v{s}
         }
\address  {
                     Institute of Nuclear Physics,
                  Academy of Sciences of the Czech Republic,
                     CS 250 68 \v{R}e\v{z}, Czech Republic
          }

\author   {S. Pittel
           }
\address  {
                     Bartol Research Institute,
                     University of Delaware,
                     Newark, DE 19716, USA
           }
\date{\today}

\maketitle

\widetext

\begin{abstract}

Neutron-proton pairing correlations are studied within the context of two
solvable models, one based on the algebra SO(5) and the other on the
algebra SO(8). Boson-mapping techniques are applied to these models and
shown to
provide a convenient methodological tool both for solving such problems and
for gaining useful insight into general features of pairing. We first focus
on the SO(5) model, which involves generalized $T=1$ pairing. Neither
boson mean-field methods nor fermion-pair approximations are able to
describe in
detail neutron-proton pairing in this model. The analysis suggests, however,
that the boson Hamiltonian obtained from a mapping of the fermion
Hamiltonian contains
a pairing force between bosons, pointing to the importance of boson-boson (or
equivalently four-fermion) correlations with isospin $T=0$ and 
spin $S=0$. These
correlations are investigated by carrying out a second boson mapping. Closed
forms for the fermion wave functions are given in terms of the fermion-pair
operators. Similar techniques are applied -- albeit in less detail -- to the
SO(8) model, involving a competition between $T=1$ and $T=0$ pairing.
Conclusions similar to those of the SO(5) analysis are reached regarding the
importance of four-particle correlations in systems involving neutron-proton
pairing.

\end{abstract}

\vspace{0.5in}


\narrowtext

\section{Introduction}
\label{sec:intro}

\newif\iffirstfig \global\firstfigfalse

The residual interaction between the nucleons in a nucleus is expected to
contain
a strong neutron-proton pairing component on the basis of isospin-invariance
arguments. Practical manifestations of neutron-proton pairing 
have proven
elusive, however, in large part because most of the nuclei studied to date
contain a significant neutron excess. In such nuclei, the neutrons and protons
near the Fermi surface occupy different valence shells and thus cannot
effectively exploit the neutron-proton pairing interaction. Moreover, as shown
recently \cite{ELV,Dob}, even when the active neutrons and
protons occupy the same valence shell, they cannot effectively build
neutron-proton pair
correlations except in the very narrow window of $N \approx Z$.

The development of radioactive beam facilities, now taking place at many
laboratories worldwide, promises to change the experimental situation
dramatically. With these new facilities, it should be possible to access all
$N=Z$ nuclei up to $^{100}{\rm Sn}$. In many of the heavier $N \approx Z$
proton-rich
nuclei, the neutron-proton pairing degree of freedom is expected to come into
significant play.  For this reason, renewed attention is now being devoted
to the
theoretical aspects of neutron-proton pairing.  Since {\em full} shell-model
calculations are feasible only for a limited set of nuclei, approximate methods
are needed to study in detail such collective features. Historically, the method
of choice has involved a generalization of the usual BCS treatment of pairing
between like nucleons \cite{Goodman,EPSVD}. Unfortunately, this approach does
not seem able to provide a correct description of many features of
neutron-proton
pair correlations \cite{CRV}. Isospin projection seems to be a promising
avenue to
an improved theory \cite{ELV}, but up to now it has not been implemented.

In studying the effects of neutron-proton pairing, simple models can be very
useful. Several such models, containing a semi-realistic representation of the
different modes of pairing, have been constructed \cite{SO5,SO8}.  These models
allow for exact solution and simple examination of various aspects of
approximate
treatments. Furthermore, they seem to reflect many of the key features
of pairing
that show up in more realistic shell-model calculations \cite{ELV}.

In the present paper, we employ the technique of boson mappings \cite{KM} to
study pairing effects in these models. The basic idea of a boson mapping -- as
traditionally implemented -- is to map bi-fermion operators onto boson 
operators in such a way as to preserve the physics of the original 
fermion problem.  In principle, the original fermion problem 
can be completely solved within the boson
space. In realistic applications, however, the boson mapping must be combined
with approximation techniques.

Boson mappings not only provide a relatively simple methodology for 
treating the original fermion problem, but also can shed light on 
pairing correlations and on
the applicability of pair approximations at the fermion level. The bosons are
images of fermion pairs. Therefore, if the mapping does not result in a
simple description in terms of the basis bosons, there would correspondingly 
not be a simple description in terms of fermion pairs.

Boson-mapping techniques may also be useful in an extended sense, by 
providing a natural 
methodology for incorporating correlations involving more than just two
particles. In the neutron-proton pairing models to be discussed, inspection of
the boson-mapped Hamiltonian suggests a pair collectivity between the bosons
introduced in the mapping. This boson pair collectivity can then be treated 
with a second boson mapping \cite{secbosm}, whereby bi-boson operators are
mapped onto new bosons representing quartets of the original fermions. 
The original fermion problem can then be rephrased in the language of 
these new bosons. A simple description in terms of these (quartet) bosons 
would confirm the importance of the associated four-fermion correlated 
structures.

The paper is organized as follows. We first consider the SO(5) model of
monopole-isovector pairing. In Section \ref{sect5bmf}, we briefly discuss the
model and its boson mapping and then examine in detail several variants of 
boson
mean-field approximations. The applicability of fermion-pair approximations
is then studied in Section \ref{sect5fpa}. In Section \ref{so5q}, a second
boson mapping of the SO(5) model is performed and the role of bi-boson
(four-fermion) structures is investigated. We then turn to the SO(8) model with
both isoscalar and isovector degrees of freedom in Section \ref{so8m}. The same
issues are addressed as for the SO(5) model, but without as much detail.
Finally,
Section \ref{discus} summarizes the key conclusions of the work and spells out
some issues for future consideration.

\section{SO(5) model - boson mean-field methods}
\label{sect5bmf}

\subsection{The SO(5) model and its boson realization}

The SO(5) model is perhaps the simplest tool to meaningfully investigate
neutron-proton pairing. In this model, a system of $N$ nucleons occupying a 
set of degenerate single-particle orbits with total degeneracy 
$4\Omega = 2\sum (2j+1)$ interact via an isovector pairing interaction,

\begin{equation}
H=g S^{\dagger} \cdot \widetilde{S} ~,
\label{hamso5}
\end{equation}
where
\begin{equation}
S^{\dagger}_{\nu}=\frac{1}{2} \sum_{j} \hat{\jmath}
[a^{\dagger j \frac{1}{2}} a^{\dagger j \frac{1}{2}}]^{01}_{0\nu} \; \; ,
\end{equation}
and

\begin{equation}
\widetilde{S}_{\nu}=(-)^{1-\nu}S_{-\nu} ~.
\end{equation}

The Hamiltonian (\ref{hamso5}) is invariant under the group of SO(5)
tranformations generated by the three pair creation operators
$S^{\dagger}_{\nu}$, the three conjugate pair annihilation operators $S_{\nu}$,
and the three components of the isospin operator ${\cal T}$.  This makes the
analysis of the model extremely simple and it is for this reason that it
has been used recently in several studies of relevance to 
neutron-proton pairing \cite{ELV,CRV,Dob}.

A Dyson boson realization of the SO(5) algebra has been constructed in Ref.
\cite{GH}. For the isovector pairing model, the boson space is constructed in
terms of a scalar ($L=0$ $S=0$ $J=0$) isovector ($T=1$) boson $s_{\nu}$, and 
the mapping takes the form \cite{Dob}
\begin{eqnarray}
 S^{\dagger}_{\nu} &\rightarrow& (\Omega - \hat{{\cal
N}}+1)s^{\dagger}_{\nu} - \frac{1}{2} s^{\dagger} \cdot s^{\dagger}
\widetilde{s}_{\nu} ~,  \nonumber \\
\widetilde{S}_{\nu} &\rightarrow&
\widetilde{s}_{\nu} ~, \nonumber \\
{\cal T}_{\nu} &\rightarrow& \sqrt{2} [ s^{\dagger}
\widetilde{s} \, ]^{01}_{0 \nu} ~.
\label{Dyson}
\end{eqnarray}
Here, $\hat{{\cal N}} = -s^{\dagger} \cdot \widetilde{s}$ is the boson number
operator.
In these and all subsequent expressions, we use the standard notation for the
scalar product, \[ s^{\dagger} \cdot s^{\dagger} = \sum (-)^{\nu}
s^{\dagger}_{\nu} s^{\dagger}_{-\nu} \; \; . \]

\subsection{Approximate boson mean-field methods}
\label{mfboson}

Boson mappings provide an alternative technique for solving a fermion
problem. Diagonalizing the mapped Hamiltonian in the ideal boson space (where
possible) would yield all of the eigenvalues and the boson images of all
eigenvectors of the original fermion problem.  Note, however, that in the boson
space spurious states may occur. These are boson states with no counterparts in
the original fermion space that arise as a pure 
artifact of the mapping. In
physically realistic problems, where it is impossible to diagonalize the mapped
Hamiltonian in the full boson space, these spurious states cannot be readily
removed from the problem. In the SO(5) model, however, this is not the case.
Here it is possible to exactly diagonalize the mapped Hamiltonian. Furthermore,
it can be shown that the spurious states only arise
for $N>2\Omega$ and
that they all have isospin $T>2\Omega-N/2$.
Thus, by
focusing our analysis on systems with $N\leq 2\Omega$, we can be sure that
spurious boson states do not contaminate the physics of interest.

In the present study, we also treat the dynamics of the boson problem
using approximate methods. By comparing with the exact results, we can
assess the
usefulness of these approximate methods in capturing the dominant collective
dynamics in the presence of pair correlations.

We begin by considering boson
mean-field techniques. The starting point is to introduce a mean-field boson
$\gamma$ as a linear combination of the basis bosons.  The creation operator 
for the mean-field boson can thus be expressed as
\begin{equation}
\gamma^{\dagger}=
\gamma_1 s^{\dagger}_{1}+ \gamma_{-1} s^{\dagger}_{-1}
+\gamma_0 s^{\dagger}_{0} \; \; .
\label{meanboson}
\end{equation}

From this mean-field boson, several approximate variational states can be
considered.
Minimization of the energy of these variational states subject to
the relevant constraints defines possible boson mean-field approximation
procedures, each of which is the natural analog of a fermion variational
procedure. We will denote each boson variational procedure by the
corresponding standard fermion terminology.

In the boson analog of the BCS
method, for example, the variational state is
\begin{equation} 
|{\rm BSC})
\propto \exp (\eta \gamma^{\dagger}) |0) ~, \label{BCS}
\end{equation}
with
constraints  on the number of bosons (one-half the number of fermions)
\[ ({\rm BCS}| \hat{N}/2 |{\rm BCS}) = {\cal N} \]
and  on the $z$-component $T_z$ of the isospin 
\[ 
({\rm BCS}|
{\cal T}_0 |{\rm BCS}) = T_z \; \; . 
\]

The boson analog of number-projected BCS is the Hartree-Bose procedure, whereby
the variational state is of the form of a boson condensate\footnote{Throughout
the present paper, we understand the projection as performed before variation.}
\begin{equation}
|{\rm BSC},{\cal N}) \propto \gamma^{\dagger {\cal N}}|0) 
\label{BCSN}
\end{equation}
with the single constraint
\[ ({\rm BCS},{\cal N}| {\cal
T}_0 |{\rm BCS},{\cal N}) = T_z \; \; . \]

The next level of mean-field approximation we consider involves the
number-$T_z$ projected BCS variational state,
\begin{equation} 
|{\rm BSC},{\cal N} \, T_z ) 
\propto {\cal P}_{T_z} \gamma^{\dagger^{\cal N}}|0) \; \; .
\label{BCSNTz}
\end{equation}
Here, ${\cal P}_{T_z}$ projects the boson condensate onto a state with definite
$T_z$.

Finally, if we were to apply, in addition to number and $T_z$ projection, full
isospin $T$ projection, the exact SO(5) state would be realized (as long as
the probe function has a nonzero overlap with the exact state).

\subsection{Energies}

We now compare the energies arising at the 
various levels of boson mean-field
approximation with the exact ground-state energies of the SO(5) Hamiltonian
(\ref{hamso5}). We consider systems with an 
even number of nucleons only.

The exact eigenenergies for a system with ${\cal N}$ nucleon pairs are given
by
\begin{equation}
E=-g[({\cal N}-\frac{1}{2}v_s)(\Omega+\frac{3}{2}-\frac{1}{2}{\cal
N}-\frac{1}{4}v_s) -\frac{1}{2}
T(T+1)] \; \; , 
\label{eSO5} 
\end{equation} 
where $v_s$ is the singlet-pairing seniority.

We are especially interested in the ground state of the system, which is
realized
for $v_s=0$. The ground-state energy is then given by
\begin{eqnarray} 
E_{{\rm exact}}&=&-g[{\cal
N}(\Omega+\frac{3}{2}-\frac{1}{2}{\cal N}) \nonumber \\ & & \; \; \; \; \; \;
-\frac{1}{2} T_z(T_z+1)-\delta (T_z+1)] \; \; , 
\label{HSO5}
\end{eqnarray} 
where $\delta=0$ for even-even ($T=T_z$) systems and 1 for 
odd-odd ($T=T_z+1$) systems.

Applying the Dyson boson mapping (\ref{Dyson}) to the SO(5)
Hamiltonian (\ref{hamso5}) leads to the  boson Hamiltonian
\begin{equation}
H_{{\rm B}} =
-g[\hat{{\cal N}} (\Omega+1-\hat{{\cal N}})+
\frac{1}{2} s^{\dagger} \cdot s^{\dagger}
\widetilde{s} \cdot \widetilde{s}]
\; \; .
\label{HSO5B}
\end{equation}
When this Hamiltonian is used in the variational
approximations described in the preceding subsection,
only two kinds of solutions can occur.
The first, which we denote as A, has
$\gamma_0=0$; the second, denoted by B, has
$\gamma_1=\gamma_{-1}=0$, $\gamma_0=1$.\footnote{ This notation
is in correspondence with Ref.\cite{EPSVD}.}

In the BCS case, with trial state (\ref{BCS}), the solutions A and B are
degenerate for $T_z=0$, whereas only solution A applies for $T_z>0$. Note
that this is precisely what was found in the generalized BCS treatment of Ref.
{\cite{EPSVD}. The variational energy for these solutions is 
\[ E({\rm BCS}) =E_{{\rm exact}} +
g[\frac{3}{2}{\cal N}-\frac{1}{2}T_z-\delta (T_z+1)] \; \; . \]

Adding number projection via the trial state (\ref{BCSN}) leaves many 
properties
of the solution(s) unchanged.  The solutions A and B are still degenerate for
$T_z=0$.  And, for $T_z>0$,  only the solution A applies. There is a change in
the variational energy, however, which now becomes 
\begin{eqnarray} 
& &E({\rm BCS},{\cal N}) \nonumber \\
& & \; \; \; \; = E_{{\rm exact}} +g{\cal N}
[1-\frac{1}{2}\frac{T_z}{{\cal N}} (1+\frac{T_z}{{\cal N}})]-g\delta (T_z+1) 
\;\; . 
\label{EBCSN} 
\end{eqnarray} 
Note that for $T_z \rightarrow {\cal N}$, the
approximate energy at this level of approximation approaches the exact result,
as expected for a degenerate-orbit pairing model with one type of nucleon.

When imposing $T_z$ projection as well, some new features appear.
For even-even nuclei, for example, $T_z$ projection removes the degeneracy
of the solutions A and B, the solution A giving the lower energy. Thus, for
even-even nuclei, the ground state solution is of the form A with
variational energy
\begin{equation}
E({\rm BCS},{\cal N} \, T_z) = E_{{\rm
exact}} +g\frac{1}{2}{\cal N}(1-\frac{T_z}{{\cal N}})  ~.
\label{EBCSNT}
\end{equation}
Here, too, for $T_z \rightarrow {\cal N}$, the approximate result approaches 
the exact result, as clearly it should.

For odd-odd nuclei, on the other hand, the solution A disappears. Clearly, an
odd-odd system must contain at least one $T_z=0$ ($s_0$) boson,  and such
components are not present in solution A.

In fact, problems with the solution A in odd-odd nuclei already show up at the
level of number-projected BCS approximation, though not as transparently. For an
odd-odd nucleus with $T_z={\cal N}-1$, the number-projected variational energy
(see Eq. \ref{EBCSN}) is lower than the exact 
energy.\footnote{The same effect is seen in the fermion generalized BCS
calculations of Ref.\cite{EPSVD}.}

For the odd-odd $T_z=0$ nucleus, the number-projected solution is thus of
the form B and already
has good $T_z$. Furthermore, the variational
energy at this level of approximation is 
\[ E({\rm BCS},{\cal N}\, T_z=0) =E_{{\rm exact}} +g({\cal N}-1) \; \; . \] 
\begin{figure}[bt]
\epsfxsize 13.5cm
\centerline{\epsfbox{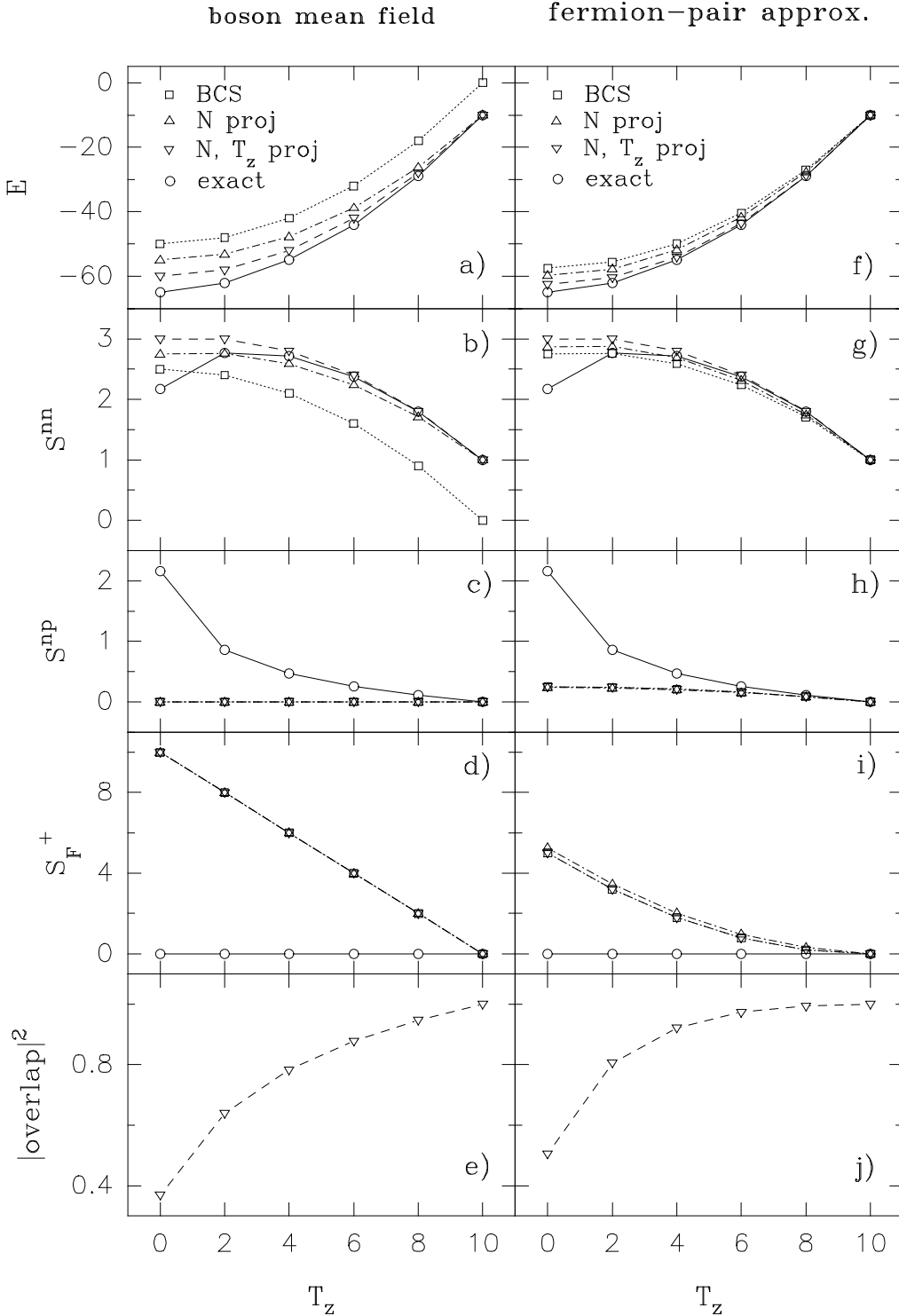}}
\caption[]{
The exact and  BCS SO(5) ground-state energies $E$
in units of $g$,
two-particle transfer strengths
$S^{{\rm nn}}$ and $S^{{\rm np}}$,
the Fermi strengths $S_{{\rm F}}^+$, and overlaps
of exact and BCS wave functions are shown
as a function
of $T_z$ for $\Omega=10$ and ${\cal N}=10$.
The results of the boson mean-field approximations
and the fermion-pair approximations are displayed
in the left and right parts of the figure, respectively.
The exact results are denoted by circles,
the BCS results by squares, the number projected BCS results
by triangles pointing up, and the number-$T_z$ projected BCS results
by triangles pointing down.
}
\label{fig1}
\end{figure}
For the odd-odd $T_z>0$ nucleus, there is no number-$T_z$ projected 
variational solution from the above-discussed class of trial functions.

A comparison of the variational energies obtained at the various levels of
approximation discussed above with the exact ground-state energies is given in
Fig. 1a. Two points are immediately evident from the figure.
First, as expected, the successive restoration of symmetries improves the
agreement with the exact results. Second, the variational energies, independent
of the specific mean-field approximation, agree quite well with the
exact values.
This can be readily understood from the structure of the boson Hamiltonian
(\ref{HSO5B}). In this Hamiltonian, the first term only depends on 
$\hat{{\cal N}}$, and its contribution to the variational energy for 
states with a given
number of bosons is not dependent on the form of the wave function.
Moreover, the
largest contribution to the variational energy from the
second term
$-\frac{1}{2}g s^{\dagger} \cdot s^{\dagger} \widetilde{s} \cdot \widetilde{s}~$
in (\ref{HSO5B}) is $~-\frac{1}{2}g {\cal N}^2~$ for all the above-discussed
solutions. As a consequence, the ground-state energy is not 
a particularly
sensitive observable for assessing the quality of the various approximate
methods, at least for this model.

\subsection{Two-particle transfer strengths}

To assess the quality of the different variational solutions, we must consider
other observables as well. In the SO(5) model, there are not many
such possibilities. The pair operators multipled by $1/\sqrt\Omega$
give the normalized two-nucleon transfer strengths of the respective
pair \cite{EDMP}. The mean values of the product of pair operators may
then be connected with the summed strengths for two-nucleon pick-up from a 
given state $i$ \cite{medtnt},
\begin{eqnarray}
S^{{\rm np}}& = &
\langle {\cal N},i |
\frac{1}{\Omega} S^{\dagger}_{0} S_{0} |{\cal N},i\rangle
\nonumber
\\
& = &
\sum_{f} |<{\cal N}-1,f |\frac{1}{\sqrt\Omega}S_0 |{\cal N},i \rangle|^2 ~,
\end{eqnarray}
and similarly for $S^{{\rm nn}}$ and $S^{{\rm pp}}$.\footnote{ These mean
values are related to the pairing gaps of the respective
pairing modes
\cite{CRV}. Their usefulness as a rough measure
of the number of the respective pairs  has also been extensively discussed
\cite{ELV,EPSVD,Dob}. }
Since two-neutron-transfer reactions provide valuable
information on nn pairing correlations \cite{BHR}, it is expected
that np transfer will analogously be sensitive to the neutron-proton-pairing
mode.
The sum of the three total transfer strengths is simply 
related to the energy of the
spin-isospin conserving Hamiltonian (\ref{hamso5}).

Exact values of the total transfer strengths for each mode can be readily
deduced
from formulae given in Refs.\cite{ELV,Dob}. For the approximate boson mean-field
methods, we find

\begin{itemize}
\item
BCS:
\end{itemize}
\[
S^{{\rm nn}} ({\rm solution~A})=
\frac{1}{2}({\cal N}+T_z)(1-\frac{{\cal N}+T_z}{2\Omega})
\; \; ,
\]
\[
S^{{\rm np}} ({\rm solution~B})=
{\cal N}(1-\frac{{\cal N}}{2\Omega})
\; \; .
\]

\begin{itemize}
\item
Number-projected BCS:
\end{itemize}
\[
S^{{\rm nn}} ({\rm solution~A})=
\frac{1}{2}({\cal N}+T_z)(1
-\frac{{\cal N}-1}{{\cal N}}\frac{{\cal N}+T_z}{2\Omega})
\; \; ,
\]
\[
S^{{\rm np}} ({\rm solution~B})=
{\cal N}(1-\frac{{\cal N}-1}{2\Omega})
\; \; .
\]

\begin{itemize}
\item
Number-$T_z$ projected BCS:
\end{itemize}
\[
S^{{\rm nn}} ({\rm solution~A})=
\frac{1}{2}({\cal N}+T_z)(1-\frac{{\cal N}+T_z-2}{2\Omega})
\; \; ,
\]
\[
S^{{\rm np}} ({\rm solution~B})=
{\cal N}(1-\frac{{\cal N}-1}{2\Omega})
\; \; .
\]
\begin{itemize} \item All methods:
\end{itemize}

\[
S^{{\rm nn}} ({\rm solution~B})=0
\; \; ,
\]
\[
S^{{\rm np}} ({\rm solution A})=0
\; \; .\]

The exact and aproximate results for the pair-transfer strengths are compared
in Fig. 1b-c. The exact and
approximate values differ considerably, much more so than the energies. For
example, for $T=0$ states, all three exact total transfer strengths are equal
\cite{ELV}. The approximate methods, however, give a zero transfer strength for
the mode that is not present in the approximate wave function. 
Thus, even though
the sum of all transfer strengths -- as mirrored in the energy -- is 
quite close
to the exact value, its composition from the various terms of the Hamiltonian
may be completely wrong.

\subsection{Fermi strengths}

Two other physically-interesting observables in the SO(5) model are the
Fermi total strengths
\[ S_{{\rm F}}^+=<{\cal T}^-{\cal T}^+> \; \; , \] \[ S_{{\rm
F}}^-=<{\cal T}^+{\cal T}^-> \; \; . \]
These two quantities are related by
the Ikeda sum rule
\[ S_{{\rm F}}^- - S_{{\rm F}}^+ = 2 T_z \; \; . \]

The exact expression for the $S_{{\rm F}}^+$ strength is
\[ S_{{\rm F}}^+({\rm
exact})=2 \delta  (T_z+1) \; \; . \]
The results for the two types of approximate solutions, A and B, are 
independent
of the specific boson mean-field method.  They are given by
\[ S_{{\rm F}}^+({\rm
solution~A})={\cal N}-T_z \; \; , \] \[ S_{{\rm F}}^+({\rm solution~B})=2 {\cal
N} \; \; . \]

A comparison of exact and approximate results is given in Fig. 1d. We see that
the approximate results may differ substantially from the exact ones, 
especially for small values of $T_z$. This again suggests that 
even if the approximate wave functions provide reasonable values for 
energies their quality may be quite poor for other more-sensitive observables.

\subsection{Overlaps}

Another measure of the quality of approximate methods is the overlap of the
approximate wave functions with the exact ones. Such a measure is not
particularly useful when considering the BCS and 
number-projected BCS
approximations, however, where the approximate wave functions are averaged over
nuclei with different number of nucleon pairs and/or 
different $T_z$.
Therefore, we calculate the overlaps only for the number-$T_z$ projected BCS
method. The results are illustrated in Fig. 1e.

We see that for small values of $T_z$, the approximate
solution has very little overlap with the exact solution. It is only for $T_z
\rightarrow {\cal N}$  that the overlap approaches one. In this limit, the
solution A (a pure nn condensate) is the exact ground state of the model system.
In the limit of symmetric nuclei, however, the approximate ground state wave
function is very bad and a more sophisticated procedure is necessary in order to
get an acceptable description of the system.

\section{SO(5) model - fermion-pair approximations}
\label{sect5fpa}

In this section, we discuss the solution of the SO(5) model in the original
fermion space. While the exact solutions and the BCS solutions have been
discussed before, the effect of number and $T_z$ projection in the BCS
approach has to our knowledge never been studied.

As noted in Subsect. \ref{mfboson}, each of the boson mean-field approximations
that we consider is the analogue of a fermion-pair approximation. The
variational wave functions associated with these fermion-pair approximations
can be readily obtained from  (\ref{BCS}), (\ref{BCSN}),
and (\ref{BCSNTz}) by replacing the collective mean-field boson
creation operator $\gamma^{\dagger}$ with a collective fermion-pair creation
operator 
\begin{equation} 
\Gamma^{\dagger}=\Gamma_1 S^{\dagger}_{1}+ \Gamma_{-1}
S^{\dagger}_{-1}+\Gamma_0 S^{\dagger}_{0} 
\label{meanpair} 
\end{equation} 
and the boson vacuum $|0)$ with the fermion vacuum $|0 \rangle$.

Despite the obvious similarities between the boson mean-field and the
corresponding fermion-pair approximations, they are not identical. Pauli
effects are accomodated very differently in the two approaches and this can
lead to differences in the results. It is only in the limit $\Omega
\rightarrow \infty$ that the corresponding bosonic and fermionic results agree.
Nevertheless, we expect the main features found in the boson mean-field 
analysis to be present in the fermion-pair methods as well.

Such a conjecture is readily confirmed for the SO(5) model under investigation.
In the fermion-pair approximations, for example, two types of solution likewise
occur. There is a solution A corresponding to
the ${\rm n \bar n -p \bar p}$ phase
($\Gamma_0=0$) and a solution B corresponding to 
the ${\rm n \bar p - p \bar n}$
phase ($\Gamma_1=\Gamma_{-1}=0$). Furthermore, as in the boson mean-field
treatment, ($i$) $T_z$ projection removes the degeneracy of the solutions A
and B, and ($ii$) both solutions are unphysical for odd-odd nuclei,
except in the
case of $T_z=0$ where the solution B is relevant.

Simple analytic expressions can be obtained in the fermion-pair
approximations for
many of the quantities discussed in Sect. \ref{sect5bmf}.
In the fermion BCS, for example, the following simple expressions obtain:
\begin{eqnarray*}
& & E({\rm BCS})
\\
& & \; \;  = E_{{\rm exact}}
+g[{\cal N}(\frac{3}{2}-\frac{3{\cal N}}{4\Omega})
-\frac{1}{2}T_z(1+\frac{T_z}{2\Omega})-\delta (T_z+1)]  ~,
\end{eqnarray*}
\[
S^{{\rm nn}} ({\rm solution~A})=
\frac{1}{2}({\cal N}+T_z)(1-\frac{{\cal N}+T_z}{2\Omega}+\frac{{\cal
N}+T_z}{2\Omega^2})  ~,
\]
\[
S^{{\rm nn}} ({\rm solution~B})=\frac{{\cal N}^2}{4\Omega^2}  ~,
\]
\[
S^{{\rm np}} ({\rm solution~A})=
\frac{{\cal N}^2-T_z^2}{4\Omega^2}  ~,
\]
\[
S^{{\rm np}} ({\rm solution~B})=
{\cal N}(1-\frac{{\cal N}}{2\Omega}+
\frac{{\cal N}}{4\Omega^2})    ~,
\]
\[
S_{{\rm F}}^+({\rm solution~A})={\cal N}-T_z
-\frac{{\cal N}^2-T_z^2}{2\Omega}   ~,
\]
\[
S_{{\rm F}}^+({\rm solution~B})=2{\cal N}(1-\frac{{\cal N}}{2\Omega})
\; \; .
\]

For the number and number-$T_z$ projected BCS, we have not obtained closed
expressions for all of the quantities of interest, even though it is likely that
they too can be derived. For those cases, the
results we present have been obtained numerically. 
The one quantity for
which we have obtained an analytic expression is the energy of the number-$T_z$
projected state,
\begin{equation}
E(BCS,{\cal N} T_z) ~=~ E_{{\rm exact}} + \frac{g}{2} {\cal N} (1-
\frac{T_z}{\cal N} - \frac{{\cal N}^2 - T_z^2}{2{\cal N}\Omega} ) ~.
\label{FBCSNTz}
\end{equation}

In the right half of Fig.\ref{fig1}, the results 
of the approximate
fermion-pair methods are shown next to their corresponding boson mean-field
results. The fermionic results are in general closer to the exact ones
than their
bosonic counterparts. Nevertheless, the main deficiencies found in the bosonic
analysis persist when working directly in the fermion space. Most notable
are the
large disrepancies in the Fermi and two-particle transfer strengths and
in the overlaps
between the approximate and exact wave functions.

Closer inspection of the results indicates that even for the relatively
well-reproduced ground-state energies the approximate methods do not
capture some
important details. An example is the double binding energy
difference \cite{ZCB},
\begin{eqnarray*}
\delta V_{{\rm np}}(N,Z)&=& \frac{1}{4}\{[B(N,Z)-B(N-2,Z)] \\ &
& \; \; \; \; -[B(N,Z-2)-B(N-2,Z-2)]\} ~. 
\end{eqnarray*}
For even-even nuclei
in this model, the exact expression for this quantity is
\begin{eqnarray*} 
\delta V_{{\rm np}}(N,Z)&=&
    -\frac{1}{4} g  \;  \; \; \;  ,~ {\rm for} \;  N=Z
\nonumber  \\
 &=& ~0    \; \; \; \; \;  \; \; \; \, ,~  {\rm otherwise}  \; \; .
\end{eqnarray*}
The jump of $\delta V_{{\rm np}}(N,Z)$ at the $N=Z$ line, which is a persistent
feature of experimental data \cite{IWB}, may be used to isolate the Wigner term
in the binding energy \cite{Satula}.\footnote{Of course, the actual Wigner term
may not have its main contribution coming from isovector pairing, as it does in
the SO(5) model.}

When $\delta V_{{\rm np}}(N,Z)$ is calculated using
the fermion number-$T_z$ projected BCS method (\ref{FBCSNTz}) -- the
approximation that yields the best reproduction of the exact energies --  one
obtains \cite{Dob} 
\[ 
\delta V_{{\rm np}}(N,Z)=
   - \frac{1}{4\Omega} ~ g   ~,
\]
irrespective of whether $N=Z$ or not. Clearly, the physics of
the jump at $N=Z$ is rather subtle and its correct
description needs more sophisticated approximate methods than BCS or any of its
variants.

Apparently, the standard fermion-pair approximations do not allow for
the coexistence of like-particle and neutron-proton pairs {\em in the
case of an isospin-conserving Hamiltonian}. This
results in a very small value of the two-nucleon transfer strength
for the mode not present in the wave function. As shown recently
\cite{CRV}, all three pairing modes can coexist, however, when the
isospin symmetry of the Hamiltonian is violated. This led to a proposal
to introduce an isospin-breaking Hamiltonian, chosen to reproduce at BCS level
the various pairing gaps in symmetric nuclei, and to use it in BCS treatments
of all nuclei.

We question such a prescription. One can
rephrase the prescription of Ref.\cite{CRV} and look for the
parameters $\Gamma_{\nu}$ in the general pair creation operator (\ref{meanpair})
that in $T=0$ nuclei give appropriate two-nucleon transfer 
strengths (or pairing
gaps) for all three modes. The condition within the SO(5) model
is that all three strengths should be equal. Working in the
number-$T_z$ projected BCS
approximation and focusing on the case $\Omega=10$, ${\cal N}=10$, we find 
that this equality can be achieved by choosing 
$\Gamma_{1}^2=\Gamma_{-1}^2 =0.2593$, and $\Gamma_{0}^2=0.4815$. 
When we then calculate the 
ground-state energy and the
Fermi strength with this same choice of structure coefficients (for the same
$T=0$ nucleus), we obtain the results -10.71 and 108.6, respectively. These are
to be compared with the exact results of -65.0 and 0, respectively, calculated
using the isospin-invariant Hamiltonian. Clearly, the proposed prescription
cannot describe in a unified way the various observables of interest.

We can understand these results as follows. The fermion-pair state just
described, namely the one that yields equal pair-transfer strengths for
all three
modes, is in fact very close to the exact state with {\em maximal} isospin
$T_{{\rm max}}={\cal N}$,
\begin{equation} 
|BCS:  {\cal N} \, T_{{\rm max}} \, T_z \rangle \propto {\cal
P}_{T_z} (\frac{1}{2} S^{\dagger}_{1}+ \frac{1}{2}
S^{\dagger}_{-1}+\frac{1}{\sqrt 2} S^{\dagger}_{0}) ^{\cal N}|0 \rangle ~.
\label{maxtstate} 
\end{equation}
This suggests that the isospin-breaking procedure proposed in Ref.\cite{CRV}
produces high-isospin admixtures in the ground state that are too large.
And this makes questionable its usefulness in describing the ground state of an
isospin-conserving Hamiltonian when $T_z \approx 0$.

It is useful to summarize here the principal findings up to this point in the
analysis. It has been argued many times that the (generalized) BCS method cannot
properly describe neutron-proton pairing in the SO(5) model \cite{CRV}. The
present results show that this conclusion does not change when number and $T_z$
projection are switched on (with variation after projection). The
deficiencies of
these standard fermionic methods for treating pairing are clearly seen in the
analysis. And, furthermore, they can be alternatively seen within the context 
of analogous mean-field boson methods applied following a boson mapping 
of the model.

\section{SO(5) model - Beyond fermion-pair correlations}
\label{so5q}

Since fermion-pair approximations and the corresponding boson mean-field
methods show deficiencies when applied to the ground state of the isovector
neutron-proton pairing SO(5) model, we need to consider more sophisticated
procedures.

\subsection{BCS for boson Hamiltonian}

The boson-mapped SO(5) Hamiltonian (\ref{HSO5B}) contains, in addition to the
dominant linear term, an attractive boson-boson pairing interaction. Two bosons
only interact when in a $J=0$ $T=0$ scalar-isoscalar state. A natural first
approach to consider for a system dominated by boson pairing is boson BCS
approximation \cite {Ring}.

In this approach, a variational boson wave function of the form\footnote{We
limit our discussion here to even-even systems.}  
\begin{equation}
|\Phi ) \propto \exp (\eta s^{\dagger}_1 +
\zeta s^{\dagger} \cdot s^{\dagger}) |0)
\; \;
\label{bosonBCS}
\end{equation}
is considered.
A generalized Bogolyubov transformation,
\begin{equation}
\sigma^{\dagger}_{\nu} =
u s^{\dagger}_{\nu}-v \widetilde{s}_{\nu} -x \delta_{1 \nu} ~,
\label{bosonbogo}
\end{equation}
is then introduced, with the constraint
\[
u^2-v^2=1
\; \; .
\]
The state (\ref{bosonBCS}) is the vacuum for quasiboson operators
(\ref{bosonbogo}) if the relations
\begin{eqnarray*}
2u\zeta + v &=& 0 ~,\\
u\eta -x &=& 0
\end{eqnarray*}
are satisfied.\footnote{In fact, the wave function (\ref{BCS}) of
Subsect. \ref{mfboson} can be considered in an analogous way with
$v=0$. We denote the method in the present subsection as
the boson BCS  to distinguish it from the BCS (\ref{BCS}) in the mean-field
boson
approximation. In the latter, the BCS bra state is a boson image of the
fermion BCS state.}
Constraints on the average number of pairs 
${\cal N}$ and the 
average
value of the isospin $T_z$ give
\begin{eqnarray*}
v^2 &=& \frac{{\cal N} - T_z}{2T_z+3} ~,\\
x^2 &=& T_z
\; \; .
\end{eqnarray*}
The energy of the state (\ref{bosonBCS}) is then easily calculated. We
present explicitly the result for $T_z=0$ only, for which the energy is
\begin{equation}
E({\rm BCS}) = E_{{\rm exact}}
+{\cal N}(1+\frac{1}{3}{\cal N})
\; \; .
\end{equation}

An illustrative comparison of the exact and approximate boson BCS energies is
given in the upper
part of Fig. \ref{fig2}. We see that the boson BCS theory cannot
explain the energies satisfactorily.

Its failure can be traced to a very large dispersion in the pair number
contained in the wave function (\ref{bosonBCS}). For the model under 
discussion, the dispersion is given by
\begin{equation}
(\Delta {\cal N} )^2 = (\Phi | \hat{{\cal
N}}^2 | \Phi)-{\cal N}^2=\frac
{({\cal N}-T_z)({\cal N}+T_z+3)(8T_z+6)}
{(2T_z+3)^2}+T_z ~.
\label{dispso5}
\end{equation} 
The results are displayed in the lower part of Fig. \ref{fig2}. 
As is clearly evident, the dispersion  
is quite large, especially for small values of $T_z$.

\begin{figure}[htb]
\epsfysize 10.cm
\centerline{\epsfbox{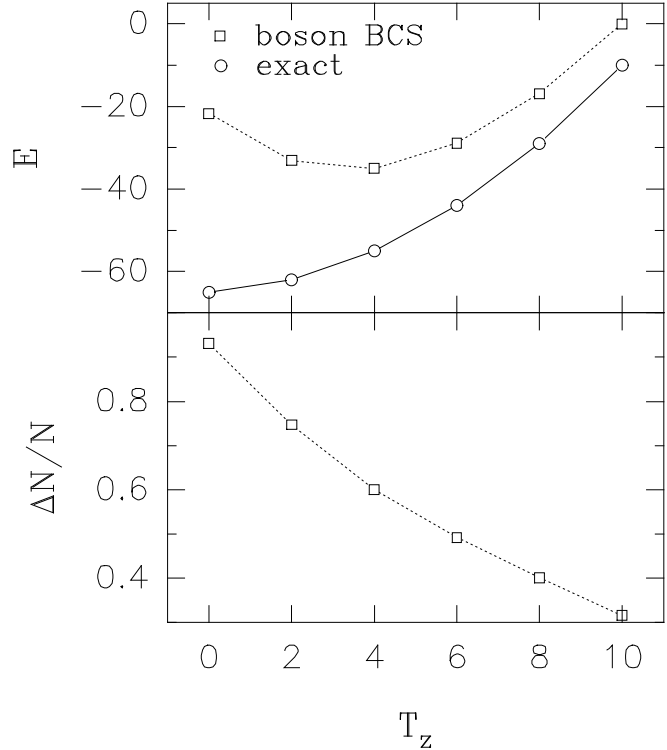}}
\caption[]{
The exact and boson BCS SO(5) ground-state energies
in units of $g$ are shown in the
upper part of the figure as a function
of $T_z$ for $\Omega=10$ and ${\cal N}=10$.
The pair-number dispersion in the boson BCS state is given in the
lower part.
}
\label{fig2}
\end{figure}

Some understanding of this result can be obtained by focussing on the case
of $T_z$=0.  
Then, Eq.(\ref{dispso5}) is a special case of a
formula for the dispersion associated
with the BCS 
wave function (\ref{bosonBCS}) for $\eta=0$ and for
pairing in a single level of degeneracy $2\Omega$: 
\begin{equation}
(\Delta {\cal N} )^2 ={\cal
N}^2/\Omega+2{\cal N} \; \; .
\label{disp1}
\end{equation}
In the case under discussion of an isovector $s$ boson, $2\Omega=3$ and
\begin{equation} 
(\Delta {\cal N} )^2 =\frac{2}{3}{\cal N} ({\cal N}+3)
\; \; . 
\end{equation}
Because of the inverse dependence on $\Omega$ of the dominant quadratic term in
(\ref{disp1}), the dispersion is quite large at $T_z$=0.\footnote{Note that
in boson BCS theory, as contrasted to fermion BCS theory, the linear and
quadratic contributions add coherently, giving a further reason for the large 
dispersion.} 

Thus, due primarily to the small dimensionality of the $s$-boson space ($d=3$),
boson BCS is not an acceptable approach for treating boson-boson pairing
correlations in this model. It should be noted, however, that when number and
$T_z$ projection are applied to (\ref{bosonBCS}) the exact ground-state results
are obtained.

\subsection{A second boson mapping}
\label{so5sbm}

As we saw in Section 2, a boson mapping followed by a mean-field treatment
of the
resulting Hamiltonian is an alternative to a  BCS treatment of the original
problem. Thus, we will now consider the possibility of implementing a second
mapping of the problem, from the system of
$s_{\nu}$ bosons (representing fermion
pairs) to a new system of bosons that represent fermion quartets.  We will 
first develop and apply the ideas to systems 
with $T=0$ and then subsequently (for
reasons to be clarified later) to systems with $T \neq 0$.

\subsubsection{$T=0$ case}

In the boson-mapped SO(5) Hamiltonian (\ref{HSO5B}) only the operators
\begin{equation} 
s^{\dagger} \cdot s^{\dagger} \; \; ~, \;\; \; \widetilde{s}
\cdot \widetilde{s} \; \;  ~, \; \; \; s^{\dagger} \cdot\widetilde{s} \; \;
\label{oper} 
\end{equation}
appear. This set of operators closes under commutation,

\begin{eqnarray}
[s^{\dagger} \cdot\widetilde{s},s^{\dagger} \cdot s^{\dagger}] &=& -2s^{\dagger}
\cdot s^{\dagger} ~, \nonumber \\ 
{\protect [}s^{\dagger}
\cdot\widetilde{s},\widetilde{s} \cdot \widetilde{s}] &=& 2\widetilde{s} \cdot
\widetilde{s} ~, \nonumber \\ 
{\protect [}s^{\dagger} \cdot
s^{\dagger},\widetilde{s} \cdot \widetilde{s}] &=& -4s^{\dagger}
\cdot\widetilde{s}+6 \; \; , 
\label{comm} 
\end{eqnarray}
generating the algebra O(2,1).

In the second boson mapping, the boson space built up from the three $s_{\nu}$
bosons is mapped onto a new space built up in terms of a single $J=0$ $T=0$
boson, which we denote $t$. Whereas the $s_{\nu}$ bosons represented pairs
of the
original fermions, these new bosons represent pairs of $s$ bosons, or
equivalently quartets of the original fermions.
It should be clear at this point
that the mapping just outlined can only give information about states with 
total $T=0$.

In more detail, the mapping is constructed such that the commutation
algebra of bi-boson operators (\ref{oper}) in the $s$-boson space are preserved
in the $t$-boson space. The Dyson realization of the mapping for the O(2,1)
algebra is
\begin{eqnarray} 
s^{\dagger} \cdot s^{\dagger} & \rightarrow &
6t^{\dagger}+4t^{\dagger}t^{\dagger}t  ~,\nonumber \\
\widetilde{s} \cdot
\widetilde{s} & \rightarrow & t \ ~,\nonumber \\
s^{\dagger} \cdot\widetilde{s} &
\rightarrow & -2t^{\dagger}t \; \; . 
\label{Dyso21} 
\end{eqnarray} 
Here, $t^{\dagger}$ and $t$ denote the creation and 
annihilation operators of the boson $t$.

We obtain after the second boson mapping a U(1) Hamiltonian
\begin{equation} 
H_{{\rm BB}}=-2g(\Omega+\frac{3}{2}-t^{\dagger}t)t^{\dagger}t \; \; . 
\label{U1H}
\end{equation}

The eigenfunctions and eigenvalues of this Hamiltonian can be simply obtained.
The eigenfunctions are condensates of $t$ bosons, $|t^n))$ \footnote{We use a
double bracket $))$ to denote a state in the $t$-boson space.} and the
corresponding eigenenergies are
\begin{equation} E=-2g((\Omega+\frac{3}{2}-n)n \;
\; . \end {equation} These eigenenergies are exactly equal to the ground-state
energies (\ref{HSO5}) of the original fermion SO(5) Hamiltonian
(\ref{hamso5}) if
the obvious relation $n=\frac{1}{2}{\cal N}$ is invoked.

There are several points that should be made before proceeding to the 
$T \neq 0$ case:
\begin{itemize}
\item The exact eigenstate $|t^n))$ is in the
form of a Hartree-Bose
approximate solution for the Hamiltonian $H_{\rm BB}$. Equivalently, following
the terminology of Sect. \ref{mfboson}, we can think of it as representing the
number-projected BCS approximation of $H_{\rm B}$. Thus, we have confirmed that
BCS approximation of the mapped Hamiltonian $H_{\rm B}$ is an acceptable way to
describe the full correlation structure of the problem, {\em as long as
number  projection is included}. 
\item Clearly, the solution $|t^n))$ involves
four-fermion correlations since the boson $t$ itself represents a correlated
quartet. Thus, we have confirmed the essential role played by four-body
correlations in systems involving both like-particle (nn and pp) pairing
correlations and np pairing correlations. 
\item Lastly, we have demonstrated,
through the example studied here, the possible usefulness of iterative boson
mappings in accomodating many-particle correlation structures. Considering the
difficulty in building quasiparticle methods that can treat many-particle (e.g.,
quartet) correlations directly at the original fermion level, we feel that this
could be an important conclusion. 
\end{itemize} Of course, we should bear
in mind that all of the above points have been demonstrated so 
far for $T=0$ systems only.

\subsubsection{$T \neq 0$ case}
\label{so5Tneq0}

The boson space introduced in the previous subsection is constructed entirely 
in terms of the $J=0$ $T=0$ $t$ boson. Thus, as noted earlier, the Hamiltonian
(\ref{U1H}) derived using the mapping (\ref{Dyso21}) to that space can only
provide information on even-$\cal{N}$ $J=0$ $T=0$ states.

To make possible a more complete analysis, we have to extend the set of
operators (\ref{oper}) considered for the second boson mapping. If we
want to also include nuclei with ${\cal N}={\rm odd}$ and/or $T_z \neq 0$,
we have to add to the set of operators (\ref{oper}) the creation and
annihilation operators of the $s$ boson, \begin{equation}
s^{\dagger} \; \;  ~, \; \;
\widetilde{s}
\; \; .
\label{oper2}
\end{equation}

We should thus consider, in addition to (\ref{comm}), the commutation
relations
\begin{eqnarray} 
[\widetilde{s}_{\nu},s^{\dagger} \cdot s^{\dagger}]
&=& -2s^{\dagger}_{\nu}  ~, \nonumber \\
{\protect [}\widetilde{s}_{\nu},\widetilde{s} \cdot \widetilde{s}] &=& 0
~, \nonumber \\
{\protect [}\widetilde{s}_{\nu} ,s^{\dagger} \cdot \widetilde{s}]
 &=&
-\widetilde{s}_{\nu}  ~,
\label{comm2}
\end{eqnarray}
and their hermitian conjugates.

The algebra of the operators (\ref{oper}) and (\ref{oper2}) have their Dyson
boson realization on the space formed by the scalar-isoscalar boson $t$
(representing, as before, fermion quartets) and the scalar-isovector boson
$\sigma_{\nu}$ (representing fermion pairs).\footnote{This procedure
is analogous
to the familiar boson-fermion mappings. In those mappings, introduced to
simultaneously describe systems with an even number and an odd number of
fermions, the set of bi-fermion and fermion operators 
is mapped onto a space
formed by boson and quasi-fermion operators.}

The Dyson realization of this extended algebra is
\begin{eqnarray}
s^{\dagger}
\cdot s^{\dagger} & \rightarrow & 6t^{\dagger}+4t^{\dagger}t^{\dagger}t
+\sigma^{\dagger} \cdot \sigma^{\dagger} -4t^{\dagger}\sigma^{\dagger} \cdot
\widetilde{\sigma} ~,\nonumber \\
\widetilde{s} \cdot \widetilde{s} & \rightarrow
& t \  ~,\nonumber \\
s^{\dagger} \cdot\widetilde{s} & \rightarrow &
-2t^{\dagger}t +\sigma^{\dagger} \cdot \widetilde{\sigma} ~, \nonumber \\
s^{\dagger}_{\nu} & \rightarrow &
\sigma^{\dagger}_{\nu}-2t^{\dagger}\widetilde{\sigma}_{\nu}  ~, \nonumber \\
\widetilde{s}_{\nu} & \rightarrow & \widetilde{\sigma}_{\nu} \; \; .
\label{Dyso212}
\end{eqnarray}

The ideal space formed by the $t$ and $\sigma$ bosons is
larger than the original $s$-boson space.
Nonphysical (spurious) states are introduced by the second boson mapping, in
addition to those already introduced by the first boson mapping.
Full diagonalization of the mapped Hamiltonian in the ideal space
separates physical states from spurious states, however, providing eigenvalues
and wave function images of the original problem.

Applying the second boson mapping (\ref{Dyso212}) to the boson Hamiltonian
(\ref{HSO5B}), we obtain
\begin{eqnarray}
H_{{\rm BB}}&=&-g[\hat{{\cal N}} (\Omega+1-\hat{{\cal N}})
+3t^{\dagger}t+2t^{\dagger}t^{\dagger}tt
\nonumber
\\
& & \; \; \; \; \; \;
-2t^{\dagger}t\sigma^{\dagger}\cdot \widetilde{\sigma}
+\frac{1}{2}\sigma^{\dagger} \cdot \sigma^{\dagger}t]
\; \; ,
\label{U1H2}
\end{eqnarray}
where we have used the notation
\[
\hat{{\cal N}}=2t^{\dagger}t -\sigma^{\dagger} \cdot \widetilde{\sigma}
\; \;
\]
for the total number of nucleon pairs in the system.  Note that the Hamiltonian
(\ref{U1H2}), when restricted to states built from $t$ bosons, 
correctly reduces
to the $T=0$ Hamiltonian (\ref{U1H}).

Though the Hamiltonian (\ref{U1H2}) is nonhermitian (a general feature of
Dyson mappings) with non-zero nondiagonal elements, we can nevertheless easily
determine its eigenenergies. In the basis with good isospin $T$,
\[ 
|t^{\frac{1}{2}({\cal N}-n_{\sigma})} ~;~
\sigma^{n_\sigma} T)) \; \; , \; \; n_\sigma
\geq T \; \; , 
\]
the Hamiltonian matrix for (\ref{U1H2}) has all zero elements
below the main diagonal. Therefore, the diagonal matrix elements coincide with
its eigenvalues, which we find to be
\[ 
-g[{\cal N}(\Omega+\frac{3}{2}-\frac{1}{2}{\cal N})
-\frac{1}{2} n_{\sigma}(n_{\sigma}+1)]
\; \; .
\]
The lowest eigenvalue, with $n_\sigma=T$, corresponds to the physical
state.\footnote{For ${\cal N}>\Omega$, the condition
$T\leq2\Omega-{\cal N}$ must also be taken into account.}
The other eigenstates are nonphysical.
The energies of the physical solutions are in precise
agreement with the exact energies (\ref{eSO5}) of the $v_s=0$
states, as they should be.

The corresponding left
eigenvector for the physical $n_{\sigma}=T$ state is readily found to be
\begin{equation} 
(( t^{\frac{1}{2}({\cal N}-T)}; \sigma^{T} T | \; \; .
\label{lefteigv} 
\end{equation}
Note further that the exact left eigenstate is a product of a
$t$-boson condensate and a
$\sigma$-boson condensate, precisely the form of a
coupled Hartree-Bose solution.

From the above remarks, we see that all of the conclusions given at the end of
the preceding subsection (for $T=0$ states) carry over to the more general case.
Namely, ({\em i}) number-projected boson BCS 
approximation following the first
boson mapping is an appropriate method for incorporating the correlations
contained in the SO(5) model, 
({\em ii}) incorporating four-particle correlations within the context 
of alpha-like clusters is the key to describing the structure
of the system, and ({\em iii}) iterative boson mappings are an attractive means
of accomodating the various correlations contained in the model.

\subsection{Wave function in terms of fermion-pair operators}
\label{wfso5}

In the previous subsection, we obtained an analytic form for the 
left eigenstates of the SO(5) model in the $t-\sigma$ space of 
the second boson mapping. For this
particular problem, we can invert the procedure to obtain the 
corresponding exact
fermion wave functions.

The key to the procedure is to focus on the left eigenstates (\ref{lefteigv}) 
of the second boson Hamiltonian. As can be seen from the mapping equations
(\ref{Dyson}, \ref{Dyso21} and \ref{Dyso212}),
there is a simple chain of relations that take us from the left fermion
eigenstates to the left second boson eigenstates, namely
\begin{eqnarray} 
\widetilde{S} \cdot \widetilde{S} &
\rightarrow \widetilde{s} \cdot \widetilde{s} & \rightarrow t ~, \nonumber \\
\protect{\widetilde{S}_{\nu}} &\rightarrow \widetilde{s}_{\nu} &
\rightarrow \widetilde{\sigma}_{\nu} \; \; . 
\label{sssst} 
\end{eqnarray}

We can therefore write the left fermion eigenstate
that leads to (\ref{lefteigv})
as\footnote{
Note that the
normalization of the fermion state cannot be simply deduced from the
boson state, so that only a proportionality relation is given.}
\[
\langle {\cal N} T| \propto
\langle T T| (\widetilde{S} \cdot \widetilde{S})
^{\frac{1}{2}({\cal N}-T)}
\; \; .
\]

Since at the fermion level the right and left eigenstates are conjugates
to one another, we can now write the right eigenstate as
\[
|{\cal N} T \rangle \propto
(\widetilde{S} \cdot \widetilde{S})
^{\frac{1}{2}({\cal N}-T)} |T T \rangle
\; \; .
\]

Finally, making use of the relation (\ref{maxtstate}), we obtain for the exact
SO(5) state vectors in terms of fermion pairs
\begin{eqnarray}
& &| {\cal N} T T_z \rangle
\nonumber
\\
& & \propto
{\cal P}_{T_z}
(S^{\dagger} \cdot S^{\dagger})^ {\frac{1}{2}({\cal N}-T)}
(\frac{1}{2} S^{\dagger}_{1}+
\frac{1}{2} S^{\dagger}_{-1}+\frac{1}{\sqrt 2} S^{\dagger}_{0})
^{T} |0 \rangle
\; .
\end{eqnarray}
A special case of this relation for the ground states of even-even 
nuclei (with $T=T_z$) and odd-odd nuclei (with $T=T_z+1$) was given 
in Ref.\cite{Dob}. Note that the exact wave functions involve a
condensate of $S^{\dagger} \cdot S^{\dagger}$ $J$=0 $T$=0 quartets and an 
isospin-stretched condensate of $J$=0 $T$=1 pairs projected to good 
$T_z$.

The above considerations reconfirm what was demonstrated in the previous
subsections, namely that four-particle correlations are an essential ingredient
for a correct description of the eigenstates of the neutron-proton pairing SO(5)
model.

\section{SO(8) model}
\label{so8m}

\subsection{The model and its boson realization}
\label{so8mbr}

A second algebraic model that has been 
used extensively \cite{SO8,EDMP,EPSVD} to
study neutron-proton pairing correlations is one based on the algebra SO(8). 
As in the SO(5) model, neutrons and protons move in a set of degenerate
single-particle orbits of total degeneracy $4\Omega$. In this model, however,
they interact via a sum of an scalar-isovector pairing interaction {\em
and} a vector-isoscalar pairing interaction, with the Hamiltonian
taking the form\footnote{We only consider the
isospin-conserving version of the SO(8) model in this work.}

\begin{equation}
H=\frac{g(1+x)}{2} ~ S^{\dagger} \cdot \widetilde{S}+
\frac{g(1-x)}{2} ~P^{\dagger} \cdot \widetilde{P}+
g_{{\rm ph}} ~ {\cal F} \cdot {\cal F}
\; \; .
\label{hamso8}
\end{equation}
Here, the operator $S_{\nu}^{\dagger}$  creates the same $L$=0 $S$=0 $J$=0
$T$=1 isovector pair as in the SO(5) model, the operator $P^{\dagger}_{\nu}$
creates an $L$=0 $S$=1 $J$=1 $T$=0 isoscalar pair, and ${\cal F}^{\mu}_{\nu}$
is the Gamow-Teller operator.

Note that the relative importance of isoscalar and isovector pairing in the
Hamiltonian (\ref{hamso8}) is governed by a single parameter $x$, which varies
from $-1$ (pure isoscalar pairing) to $+1$ (pure isovector pairing). The last
term in the Hamiltonian, a particle-hole force in the $T=1$ $S=1$ channel, is
included to bring the Hamiltonian into closer contact with more realistic
nuclear Hamiltonians, without destroying the simplicity of the model.

The Hamiltonian (\ref{hamso8}) is invariant under the group of SO(8)
transformations generated by the three isovector pair creation operators
$S^{\dagger}_{\nu}$, their three conjugate annihilation operators $S_{\nu}$,
the three isoscalar pair creation operators $P^{\dagger}_{\mu}$, their three
conjugate annihilation operators $P_{\mu}$, the three components of the
isospin operator $T_{\nu}$,
the three components of the
spin operator ${\cal S}_{\mu}$,
and the nine components of the Gamow-Teller operator
${\cal F}^{\mu}_{\nu}$.

The Dyson boson realization of the SO(8) algebra is constructed
by mapping its bi-fermion operators onto
bosonic operators formed from the creation operators
$s^{\dagger}_{\nu}$ of a scalar-isovector boson $s$ and
$p^{\dagger}_{\mu}$ of a vector-isoscalar boson $p$,
and their conjugate annihilation operators
$\widetilde{s}_{\nu}=(-)^{1-\nu}s_{-\nu}$
and $\widetilde{p}_{\mu}=(-)^{1-\mu}p_{-\mu}$
\cite{so8map}:
\begin{eqnarray}
S^{\dagger}_{\nu} & \rightarrow &
(\Omega - \hat{{\cal N}}+1)s^{\dagger}_{\nu}
+ \frac{1}{2} (p^{\dagger} \cdot p^{\dagger} - s^{\dagger} \cdot s^{\dagger})
\widetilde{s}_{\nu} ~,
\nonumber
\\
P^{\dagger}_{\mu} & \rightarrow &
(\Omega - \hat{{\cal N}}+1)p^{\dagger}_{\mu}
+ \frac{1}{2} (s^{\dagger} \cdot s^{\dagger} - p^{\dagger} \cdot p^{\dagger})
\widetilde{p}_{\mu}  ~,
\nonumber
\\
\widetilde{S}_{\nu} & \rightarrow  & \widetilde{s}_{\nu}  ~,
\nonumber
\\
\widetilde{P}_{\mu} & \rightarrow & \widetilde{p}_{\mu}  ~,
\nonumber
\\
{\cal T}_{\nu} & \rightarrow & \sqrt{2} 
[ s^{\dagger} \widetilde{s} \, ]^{01}_{0 \nu}  ~,
\nonumber
\\
{\cal S}_{\mu} & \rightarrow & \sqrt{2} [ p^{\dagger} \widetilde{p} \, ]^{10}
_{\mu 0}  ~,
\nonumber
\\
{\cal F}^{\mu}_{\nu} & \rightarrow &
-(p^{\dagger}_{\mu} \widetilde{s}_{\nu} +
s^{\dagger}_{\nu} \widetilde{p}_{\mu})
\; \; .
\label{so8bosmap}
\end{eqnarray}
Here,
\begin{displaymath}
\hat{{\cal N}} = -( s^{\dagger} \cdot \widetilde{s} +
p^{\dagger} \cdot \widetilde{p})
\end{displaymath}
is the total boson number operator.

As in the SO(5) model, spurious states are introduced by this mapping. 
Here too they arise for $N > 2 \Omega$ \cite{so8map} and can be 
readily identified when a
boson basis with good spin and isospin is employed.

\subsection{Dynamical symmetries of the model}

For certain values of its parameters, the SO(8) Hamiltonian (\ref{hamso8})
exhibits dynamical symmetries. In such cases, the eigenvalues can be obtained
analytically. For all other parameters, analytic expressions for the 
eigenvalues cannot be obtained, and numerical diagonalization is required.

The SO$^{\rm T}$(5) dynamical symmetry limit of the model is 
realized when $x=1$
and $g_{{\rm ph}}=0$. In this case,
the Hamiltonian reduces to (\ref{hamso5}) and
its ground-state energy is given analytically by (\ref{HSO5}).

The SO$^{\rm S}$(5) limit is realized for the parameters $x=-1$ and $g_{{\rm
ph}}=0$. In this case, the ground-state energy is given by
\begin{equation} 
E=-g[({\cal N}-T_z)
(\Omega+\frac{3}{2}-\frac{1}{2}{\cal N} -\frac{1}{2} T_z)-\delta]
\; \; . 
\label{egsSOs5} 
\end{equation}

The third dynamical symmetry arises when $x=0$, namely when there are equal
amounts of isovector and isoscalar pairing. In this case, the
Hamiltonian (\ref{hamso8}) has an SU(4) dynamical symmetry and its
exact eigenvalues can be written as
\begin{eqnarray} 
E & = &-\frac{1}{4}g[2{\cal N}(\Omega+3)-{\cal N}^2 -
\lambda(\lambda+4)] \nonumber \\ & &
+g_{{\rm ph}}[\lambda(\lambda+4)-S(S+1)-T(T+1)] \;
\; , 
\label{esu4} 
\end{eqnarray}
with $\lambda$ the usual SU(4) label.

The ground-state solution in this symmetry limit arises when
$\lambda=T_z+\delta$, and the corresponding energy is
\begin{eqnarray} 
E & = &-\frac{1}{4}g[2{\cal N}(\Omega+3)-{\cal N}^2 -
T_z(T_z+4)-\delta(2T_z+5)] \nonumber \\ 
& & +3g_{{\rm ph}}(T_z+\delta) \; \; .
\label{egssu4} 
\end{eqnarray}

\subsection{The first boson-mapped Hamiltonian}
\label{FBMH}

Mapping the fermion Hamiltonian (\ref{hamso8}) onto the space of $s$ and $p$
bosons leads to the boson Hamiltonian
\begin{eqnarray}
H_{\rm B}&=& -\frac{g(1+x)}{2}[(\Omega+1-\hat{{\cal
N}})s^{\dagger} \cdot \widetilde{s} +\frac{1}{2}(p^{\dagger} \cdot p^{\dagger}
- s^{\dagger} \cdot s^{\dagger})
\widetilde{s} \cdot \widetilde{s}
)] \nonumber
\\
& & -\frac{g(1-x)}{2}[(\Omega+1-\hat{{\cal N}})p^{\dagger} \cdot
\widetilde{p}
+\frac{1}{2}(s^{\dagger} \cdot s^{\dagger}-
p^{\dagger} \cdot p^{\dagger})
\widetilde{p} \cdot \widetilde{p}
)]
\nonumber
\\
& & +g_{{\rm ph}}[p^{\dagger} \cdot p^{\dagger}\widetilde{s} \cdot
\widetilde{s} +
s^{\dagger} \cdot s^{\dagger}
\widetilde{p} \cdot \widetilde{p}
+2 s^{\dagger} \cdot\widetilde{s}p^{\dagger} \cdot\widetilde{p}
+3\hat{{\cal N}}]
\; \; .
\label{so8bosham}
\end{eqnarray}
Diagonalization of the boson Hamiltonian (\ref{so8bosham}) is
straightforward and
represents an alternative method for exactly solving the SO(8) model.

\subsection{Boson mean-field and fermion-pair approximations}
\label{bmfso8}

It is of interest to consider approximate solutions for the ground state of 
this model as well. Here, too, the natural 
approximations to look at first are those
based either on mean-field boson methods or on the analogous fermion-pair
approximations.

We will not discuss in detail either the methods or the conclusions that emerge
from these approximate treatments, since they parallel so closely the 
discussion for the SO(5) model. Rather, we will just note 
some of the differences that show
up relative to the SO(5) analysis, as derived from the earlier more extensive
treatment of the SO(8) model in Ref.\cite{EPSVD}.

One of the more interesting new features that emerges is a third type of
mean-field solution. The solutions A and B, discussed in Sections
\ref{sect5bmf} and \ref{sect5fpa}, of course persist in the SO(8)
analysis. But now a third solution C also appears, corresponding to a
phase with ${\rm n \bar n -p \bar p}$ and $T=0$ ${\rm n \bar p - p \bar n}$
pairs.

As in the SO(5) analysis, there is a partial decoupling of the pairing 
phases in the mean-field solutions.  
Specifically, the $T=1$ ${\rm n \bar p - p \bar n}$
phase is absent from solutions A and C. This leads to significant
deficiencies in the mean-field description
of $T_z \approx 0$ nuclei, except when
isoscalar pairing is dominant.

\subsection{Beyond fermion-pair correlations: A second boson mapping}

The boson mean-field and fermion-pair approximations touched on in the previous
subsection incorporate fermion-pair correlations. The fact that they are unable
to describe the detailed properties of the model suggests the need for
a more sophisticated procedure that incorporates further correlations.

\subsubsection{The $T$=0 $S=0$ case}

The only operators that enter the first boson Hamiltonian
(\ref{so8bosham}) are the three operators (\ref{oper}) of the isovector boson
space and the three analogous operators 
\begin{equation} 
p^{\dagger} \cdot p^{\dagger} \; \;  ~, \;\; \; 
\widetilde{p} \cdot \widetilde{p} \; \;  ~, \; \; \; 
p^{\dagger} \cdot\widetilde{p} 
\label{operp} 
\end{equation} 
of the isoscalar
boson space. The operators (\ref{operp}) have commutation relations
analogous to (\ref{comm}). As a consequence, the Hamiltonian
(\ref{so8bosham}) exhibits an O(2,1)$\otimes$O(2,1) symmetry. Furthermore, the
two-body terms of this 
Hamiltonian are precisely in the form of
a boson pairing interaction.

As in our treatment of the SO(5) model, we choose to include the
effects of boson pairing (or equivalently four-fermion
correlations) through the use of a second boson mapping. Furthermore, we follow
the same strategy as in Sect. \ref{so5q}, first carrying out the analysis for
$T=0$ $S=0$ systems, and then generalizing.

When dealing with $T=0$ $S=0$ systems, we must map onto a boson space defined 
by two scalar-isoscalar bosons. One is the $t$ boson introduced in Subsect.
\ref{so5sbm}, which reflects the correlations of two $s$ bosons. The second,
which we denote by $q$, 
reflects the correlations of two $p$ bosons. The Dyson
realization of the O(2,1)$\otimes$O(2,1) algebra contains two parts. 
The set of
operators (\ref{oper}) of the isovector boson space map according to
(\ref{Dyso21}). The set of operators in the isoscalar space map according to the
analogous relations 
\begin{eqnarray} 
p^{\dagger} \cdot p^{\dagger} & \rightarrow
& 6q^{\dagger}+4q^{\dagger}q^{\dagger}q ~,\nonumber \\ \widetilde{p} \cdot
\widetilde{p} & \rightarrow & q ~,\ \nonumber \\
p^{\dagger} \cdot\widetilde{p} &
\rightarrow & -2q^{\dagger}q ~.
\label{Dyso21p} 
\end{eqnarray}

Applying the second boson mapping to the Hamiltonian (\ref{so8bosham}) leads to
a U(1)$\otimes$U(1) (or SU(2)) 
Hamiltonian
\begin{eqnarray}
H_{{\rm BB}}&=& -g(1+x)[(\Omega+\frac{5}{2}-\hat{{\cal N}})t^{\dagger}t
+t^{\dagger}t^{\dagger}tt -\frac{3}{2}q^{\dagger}t-q^{\dagger}q^{\dagger}qt ]
\nonumber \\ & &-g(1-x)[(\Omega+\frac{5}{2}-\hat{{\cal N}})q^{\dagger}q
+q^{\dagger}q^{\dagger}qq
-\frac{3}{2}t^{\dagger}q-t^{\dagger}t^{\dagger}tq  ~
]
\nonumber
\\
& & +g_{{\rm ph}}[
6(q^{\dagger}t+t^{\dagger}q)+4(q^{\dagger}q^{\dagger}qt+
t^{\dagger}t^{\dagger}tq)
+8t^{\dagger}tq^{\dagger}q
+3\hat{{\cal N}}]
\; \; ,
\label{so8secbosham}
\end{eqnarray}
where now $\hat{{\cal N}}$ is given by
\[
\hat{{\cal N}}=2(t^{\dagger}t+q^{\dagger}q)
\; \; .
\]

In general, the problem of the
Hamiltonian (\ref{so8secbosham}) must be solved by
numerical matrix diagonalization. In the case of the dynamical symmetry limits,
however, we can obtain simple analytic solutions.

In the SO$^{\rm T}$(5) limit, the Hamiltonian
matrix in the $|t^{\frac{1}{2}{\cal N}-n_q}q^{n_q} ))$ basis has all zeros
below the diagonal. The same considerations as in
Subsect. \ref{so5Tneq0} can therefore be applied. The diagonal elements of the
Hamiltonian matrix give directly the eigenvalues and reproduce the exact result
(\ref{eSO5}) for the $T=0$ energy when an identification $n_q=\frac{1}{4}v_{\rm
s}$ is made. The same is true for the SO$^{\rm S}$(5) limit, with the roles of
the isospin and spin and also the $t$- and $q$-bosons interchanged.

To obtain the solution in the SU(4) limit, we first impose the transformation
\begin{eqnarray}
r^{\dagger}&=&\frac{1}{\sqrt 2}(t^{\dagger}-q^{\dagger})  ~,
\nonumber
\\
w^{\dagger}&=&\frac{1}{\sqrt 2}(t^{\dagger}+q^{\dagger})  ~,
\label{su4trans}
\; \;
\end{eqnarray}
and rewrite the Hamiltonian (\ref{so8secbosham}) as
\begin{eqnarray} 
H_{{\rm BB}}&=& -g[\frac{1}{2}(\Omega+\frac{5}{2}-\hat{{\cal N}})\hat{{\cal N}}
+\frac{3}{2}(r^{\dagger}r-w^{\dagger}w)
\nonumber
\\
& & \; \; \; \; \; \; \;\; \; \; \;
+r^{\dagger}r^{\dagger}rr
+2r^{\dagger}rw^{\dagger}w
+w^{\dagger}w^{\dagger}rr
]
\nonumber
\\
& & +4g_{{\rm ph}}[3w^{\dagger}w
+w^{\dagger}w^{\dagger}ww-w^{\dagger}w^{\dagger}rr]
\; \; .
\end{eqnarray}
Working with this Hamiltonian in the basis
$|r^{\frac{1}{2}{\cal N}-n_w}w^{n_w} ))$,
we recover the exact SU(4) eigenvalues (\ref{esu4}), when an identification
$n_w=\frac{1}{2} \lambda$ is adopted.

\subsubsection{General case}

The procedure just outlined for carrying out a second boson mapping to the
$t-\sigma$ space enables investigations of ${\cal N}={\rm even}$ 
$T=0$ $S=0$ states only. For a more complete treatment, we must
appropriately extend the second boson mapping, in much the same way as we
did in Subsection \ref{so5Tneq0}. Namely, we must include, in addition to the
sets of operators (\ref{oper}), (\ref{oper2}) and (\ref{operp}), the additional
one-boson creation and annihilation operators
\begin{equation} 
p^{\dagger} \; \;  ~, \;\; \; \widetilde{p} \; \; .
\label{oper2p} 
\end{equation}

The full second boson mapping is now given by (\ref{Dyso212}) and
\begin{eqnarray} 
p^{\dagger} \cdot p^{\dagger} & \rightarrow
& 6q^{\dagger}+4q^{\dagger}q^{\dagger}q +\pi^{\dagger} \cdot \pi^{\dagger}
-4q^{\dagger}\pi^{\dagger} \cdot \widetilde{\pi}~,\nonumber \\
\widetilde{p}
\cdot \widetilde{p} & \rightarrow & q  ~, \nonumber \\
p^{\dagger}
\cdot\widetilde{p} & \rightarrow & -2q^{\dagger}q +\pi^{\dagger} \cdot
\widetilde{\pi} ~, \nonumber \\
p^{\dagger}_{\nu} & \rightarrow &
\pi^{\dagger}_{\nu}-2q^{\dagger}\widetilde{\pi}_{\nu} ~, \nonumber \\
\widetilde{p}_{\nu} & \rightarrow &
\widetilde{\pi}_{\nu}
\; \; ,
\label{Dyso212p}
\end{eqnarray}
where $\pi$ denotes the additional vector-isoscalar boson needed to complete 
the ideal boson space.

This ideal boson space -- formed from the
$t$, $\sigma$, $q$, and $\pi$ bosons -- contains an unphysical
subspace. Since the sectors $t$-$\sigma$ and $q$-$\pi$ are separated
and since we have already determined the physical states (\ref{lefteigv}) 
in the $t$-$\sigma$ sector, we find that the physical basis of the full
problem is of the form
\begin{equation} 
| t^{\frac{1}{2}({\cal N}-T-S)-n_q}
q^{n_q} ~;~ \ \sigma^{T} T; \pi^{S} S)) \; \; ~,
\; \; S + T \leq 2\Omega-\cal{N}
~. 
\label{so8pysb} 
\end{equation}

The second boson image of the
SO(8) Hamiltonian is straightforwardly obtained and
in general can be solved by matrix diagonalization. We will
discuss here only the dynamical symmetry limits, where analytical solutions 
can be obtained as in Subsection \ref{so5Tneq0}.

In the SO$^{\rm T}$(5) limit, for
example, the Hamiltonian takes the form

\begin{eqnarray}
H_{{\rm BB}}&=&
-g[(\Omega+1-\hat{{\cal N}})(2t^{\dagger}t-\sigma^{\dagger} \cdot
\widetilde{\sigma})
\nonumber
\\ & & \; \; \; \; \;
+3 t^{\dagger}t +2 t^{\dagger}t^{\dagger}tt
-2t^{\dagger}t\sigma^{\dagger} \cdot \widetilde{\sigma}
+\frac{1}{2} \sigma^{\dagger} \cdot \sigma^{\dagger} t
\nonumber
\\ & &  \; \; \; \; \;
-3 q^{\dagger}t -2 q^{\dagger}q^{\dagger}qt
+2q^{\dagger}t\pi^{\dagger} \cdot \widetilde{\pi}
-\frac{1}{2} \pi^{\dagger} \cdot \pi^{\dagger} t
]  ~,
\end{eqnarray}
with
\[
\hat{{\cal N}}=2(t^{\dagger}t+q^{\dagger}q)-
\sigma^{\dagger} \cdot \widetilde{\sigma}
-\pi^{\dagger} \cdot \widetilde{\pi}
\; \; .
\]
In the basis (\ref{so8pysb}), the exact energies (\ref{eSO5})
are obtained when the identification $2n_q+S=\frac{1}{2} v_s$ is made.
The left vectors of the basis
(\ref{so8pysb}) give the
left physical eigenvectors in this limit.

The same arguments can be applied to the SO$^{\rm S}$(5) limit. The only
difference is that we must interchange the role of the isospin with that of the
spin, and likewise the role of the $t$- and $\sigma$- bosons with that of the
$q$- and $\pi$-bosons.

In the SU(4) limit, we again apply the transformation (\ref{su4trans}),
after which the second boson Hamiltonian takes the form
\begin{eqnarray} H_{{\rm BB}}&=& -g[\frac{1}{2}(\Omega+1-\hat{{\cal
N}})\hat{{\cal N}} +3r^{\dagger}r +r^{\dagger}r^{\dagger}rr  \nonumber \\
& & +2r^{\dagger}rw^{\dagger}w +w^{\dagger}w^{\dagger}rr
-r^{\dagger}r(\sigma^{\dagger} \cdot \widetilde{\sigma}
+\pi^{\dagger} \cdot \widetilde{\pi})
\nonumber
\\
& &-w^{\dagger}r(\sigma^{\dagger} \cdot \widetilde{\sigma}
-\pi^{\dagger} \cdot \widetilde{\pi})
+\frac{{\sqrt 2}}{4}
(\sigma^{\dagger} \cdot \sigma^{\dagger}
-\pi^{\dagger} \cdot \pi^{\dagger})r
]
\nonumber
\\
& & +g_{{\rm ph}}
[6w^{\dagger}w-6r^{\dagger}r
+4w^{\dagger}w^{\dagger}ww-4w^{\dagger}w^{\dagger}rr  \nonumber \\
& &
+3\hat{{\cal N}}
-4w^{\dagger}w(\sigma^{\dagger} \cdot \widetilde{\sigma}
+\pi^{\dagger} \cdot \widetilde{\pi})
+4w^{\dagger}r(\sigma^{\dagger} \cdot \widetilde{\sigma}
-\pi^{\dagger} \cdot \widetilde{\pi})
\nonumber
\\
& &
+\frac{1}{\sqrt 2}(\sigma^{\dagger} \cdot \sigma^{\dagger}
+\pi^{\dagger} \cdot \pi^\dagger)w
-\frac{1}{\sqrt 2}(\sigma^{\dagger} \cdot \sigma^{\dagger}
-\pi^{\dagger} \cdot \pi^\dagger)r
+
2\sigma^{\dagger} \cdot \widetilde{\sigma}
\pi^{\dagger} \cdot \widetilde{\pi}
]
\; \; .
\end{eqnarray}
Again, the SU(4) energies (\ref{esu4}) are easily obtained by working in the
basis
\begin{equation}
| r^{\frac{1}{2}({\cal N}-T-S)-n_w} w^{n_w}; \
\sigma^{T} T ~;~  \pi^{S} S)) ~,
\label{so4pysb}
\end{equation}
with
$2n_w+S+T=\lambda$.
The left vectors of the basis
(\ref{so4pysb}) give the
left physical eigenvectors in the SU(4) limit.

\subsection{Wave function in terms of fermion-pair operators}

As we have just seen, the left physical
eigenvectors of the second boson-mapped space take a very simple form
in  the three dynamical symmetry limits.
As in Subsection \ref{wfso5}, we can use
this fact to find the corresponding wave functions in terms of fermion-pair
operators. To do this, we must use the chains (\ref{sssst}) as well as the
analogous $P$-sector chains 
\begin{eqnarray} 
\widetilde{P} \cdot \widetilde{P}
&\rightarrow \widetilde{p} \cdot \widetilde{p} & \rightarrow  q  ~, 
\nonumber \\
\protect{\widetilde{P}_{\mu}} &\rightarrow \widetilde{p}_{\mu} & \rightarrow
\widetilde{\pi}_{\mu}  ~.
\label{ppppq} 
\end{eqnarray}.

Proceeding  as in Subsect. \ref{wfso5}, 
we find that in the SO$^{\rm
T}$(5) limit the wave functions are given by
\begin{eqnarray}
|{\cal N} T T_z S S_z v_s \rangle & \propto & {\cal P}_{T_z}{\cal P}_{S_z}
(S^{\dagger} \cdot S^{\dagger})^ {\frac{1}{2}({\cal N}-T-\frac{1}{2}v_s)}
(P^{\dagger} \cdot P^{\dagger})^ {\frac{1}{2}(\frac{1}{2}v_s-S)}
\nonumber \\
 & & \times(\frac{1}{2} S^{\dagger}_{1}+ \frac{1}{2}
S^{\dagger}_{-1}+\frac{1}{\sqrt 2} S^{\dagger}_{0}) ^{T} (\frac{1}{2}
P^{\dagger}_{1}+ \frac{1}{2} P^{\dagger}_{-1}+\frac{1}{\sqrt 2} P^{\dagger}_{0})
^{S} |0 \rangle ~,
\end{eqnarray}
where ${\cal P}_{S_z}$ is the projection operator that picks states with
spin-projection $S_z$.

An analogous expression
holds in the SO$^{\rm S}$(5) limit with the roles of isospin and
spin and also of
$S^{\dagger}$ and $P^{\dagger}$ interchanged.

In the SU(4) limit, we must again take into account the
transformation (\ref{su4trans}). This then leads to
\begin{eqnarray}
| {\cal N} T T_z S S_z \lambda \rangle & \propto &
{\cal P}_{T_z}{\cal P}_{S_z}
(S^{\dagger} \cdot S^{\dagger}
-P^{\dagger} \cdot P^{\dagger})^ {\frac{1}{2}({\cal N}-\lambda)}
(S^{\dagger} \cdot S^{\dagger}
+P^{\dagger} \cdot P^{\dagger})^ {\frac{1}{2}(\lambda-S-T)}
\nonumber
\\
& &
\times
(\frac{1}{2} S^{\dagger}_{1}+
\frac{1}{2} S^{\dagger}_{-1}+\frac{1}{\sqrt 2} S^{\dagger}_{0})
^{T}
(\frac{1}{2} P^{\dagger}_{1}+
\frac{1}{2} P^{\dagger}_{-1}+\frac{1}{\sqrt 2} P^{\dagger}_{0})
^{S}
|0 \rangle
\; \; .
\end{eqnarray}

In the case of the ground-state solutions,
the wave functions associated with the
three dynamical symmetry limits can be unified. For even-even $N=Z$ nuclei, for
example, all three can be written as
\begin{equation} 
|{\cal N} \, T\!=\!0 \, T_z\!=\!0 \, S\!=\!0
\rangle \propto (\alpha S^{\dagger} \cdot S^{\dagger} -\beta P^{\dagger} \cdot
P^{\dagger} )^{{\cal N}/2}|0 \rangle ~. 
\label{gsform} 
\end{equation}
In the ${\rm SO}^{\rm T} (5) $ limit, $\alpha=1$ and $\beta=0$. In the
${\rm SO}^{\rm S} (5) $ limit, $\alpha=0$ and  $\beta=1$. And, in the SU(4)
limit, $\alpha=\beta=1/\surd 2$.
\footnote{Here, the normalization $\alpha^2+\beta^2=1 $ is used.}

Similarly, we can write the ground-state wave
function for even-even $N>Z$ nuclei by adding
the appropriate number of isovector
nn pairs to the $N=Z$ solution (\ref{gsform}), viz: 
\begin{eqnarray} 
& & |{\cal
N} \, T\!=\!T_z \, T_z \, S\!=\!0 \rangle \nonumber \\ & & \; \; \; \; \; \; \;
\; \propto (\alpha S^{\dagger} \cdot S^{\dagger} -\beta P^{\dagger} \cdot
P^{\dagger} )^{({\cal N}-T_z)/2}S^{\dagger T_z}_{1} |0 \rangle \; .
\label{gsform1}
\end{eqnarray}

Finally, for odd-odd nuclei, we must add to (\ref{gsform1}) either an isoscalar
or an isovector np pair, depending on the symmetry limit. In particular,
in the ${\rm SO}^{\rm T} (5) $ and  SU(4) limits, 
the ground state is given by
\begin{eqnarray} 
& & |{\cal N} \, T\!=\!T_z\!+\!1 \, T_z \, S\!=\!0 \rangle
\nonumber \\ & & \; \; \; \; \; \; \; \; \propto (\alpha S^{\dagger} \cdot
S^{\dagger}
-\beta P^{\dagger} \cdot P^{\dagger} )^{({\cal N}-T_z-1)/2}S^{\dagger
T_z}_{1} S^{\dagger}_{0} |0 \rangle ~. 
\label{xle0} 
\end{eqnarray}
Analogously, in the ${\rm SO}^{\rm S} (5) $ and  SU(4) limits, it
is given by\footnote{In
the SU(4) limit, there is a degeneracy in the ground states of odd-odd nuclei,
explaining why we give two distinct ground-state solutions.}
\begin{eqnarray} 
& & |{\cal N} \, T\!=\!T_z \, T_z \, S\!=\!1 \rangle \nonumber \\ 
& & \; \; \; \;
\; \; \; \; \propto (\alpha S^{\dagger} \cdot S^{\dagger} -\beta P^{\dagger}
\cdot P^{\dagger} )^{({\cal N}-T_z-1)/2}S^{\dagger T_z}_{1} P^{\dagger} |0
\rangle \; \; . \label{xge0} \end{eqnarray}

In the SO(8) model, the quartet creation operator that represents an alpha
cluster takes the form \cite{EDMP}\footnote{ There is a sign mistake in
Ref.\cite{EDMP}}
\[
\frac{1}{2\sqrt{3\Omega(\Omega+2)}} (S^{\dagger} \cdot S^{\dagger} -P^{\dagger}
\cdot P^{\dagger}) \; . \]
We see, therefore,  that in the SU(4) limit the ground state involves as
many alpha-correlated structures as possible.

\begin{figure}[hbt]
\epsfysize 14.cm
\centerline{\epsfbox{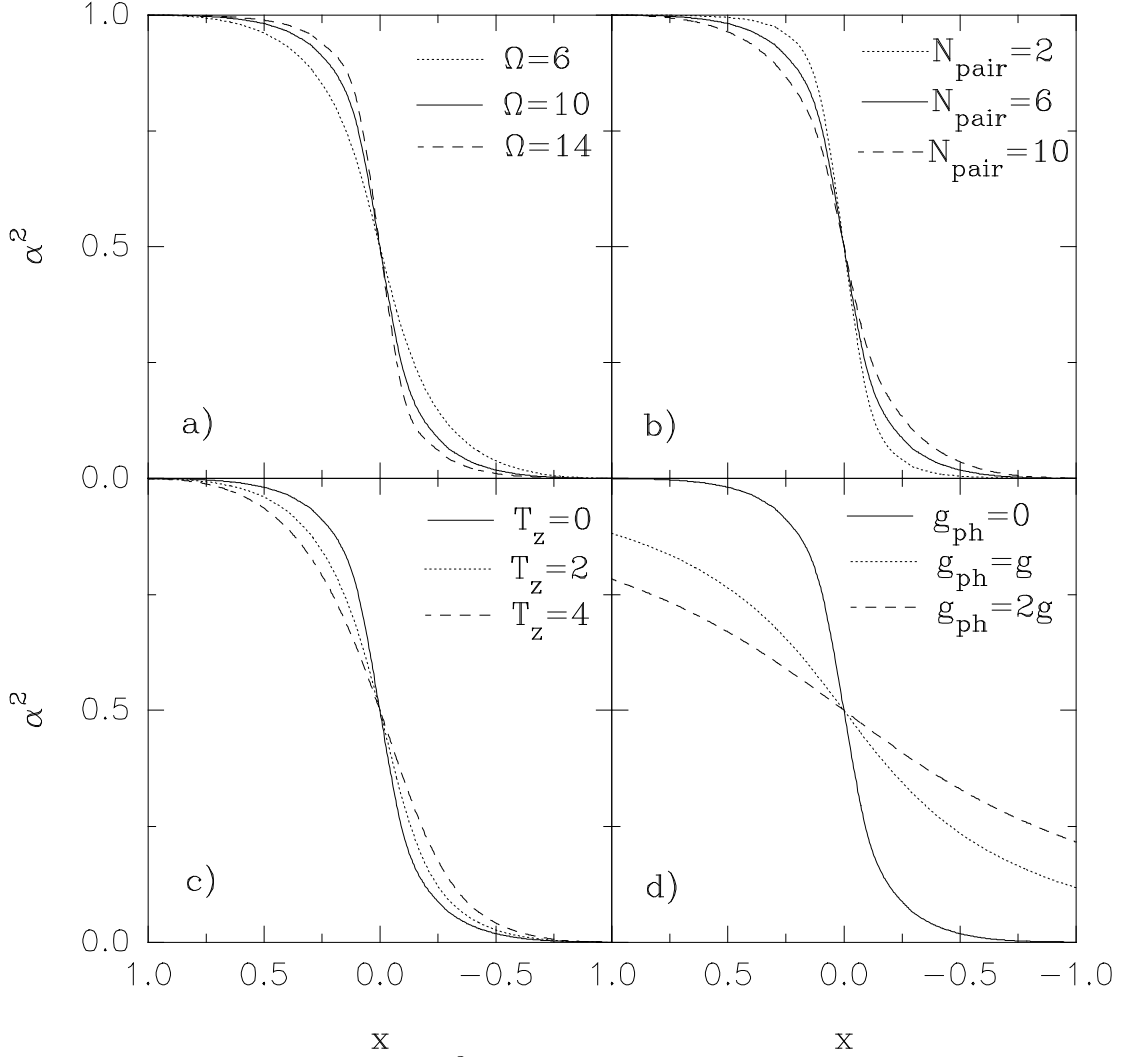}}
\caption[]{
The variational parameter $\alpha^2$ in the approximate SO(8) ground-state wave
functions (54-57) as a function of the Hamiltonian parameter $x$ that controls
the balance between isoscalar and isovector pairing. The behavior
is shown for a) various shell degeneracies $\Omega$ with fixed ${\cal N}=6$
and $T_z=0$, b) various $N_{\rm pair}={\cal N}$ with $\Omega=10$ and $T_z=0$,
c) various
$T_z$ with fixed $\Omega=10$ and ${\cal N}=T_z+6$, and d) various values of the
particle-hole strength $g_{{\rm ph}}$ in the
$T=1, S=1$ channel with fixed ${\cal
N}=6$, $T_z=0$, and $\Omega=10$.} \label{fig3} \end{figure}

Outside the dynamical symmetry limits, the above forms for the ground-state 
wave function are not exact. Nevertheless, when we consider them as
variational wave functions (with $\alpha$ and $\beta$ as
variational parameters), we find that at the minimum they have almost perfect
overlap with the exact wave functions.
Fig. \ref{fig3} illustrates the behavior of
$\alpha^2$ as a function of the parameter $x$ of the SO(8) Hamiltonian for 
a few representative cases.\footnote{ In the odd-odd case,
the form (\ref{xle0}) applies for $x \geq 0$ and
the form (\ref{xge0}) for $x \leq 0$.}

We first discuss the results in the limit $g_{{\rm ph}}=0$, as illustrated in
Fig. \ref{fig3}a-c. In this limit, we are in the pure isovector
phase at $x=1$ and in the pure isoscalar phase at $x=-1$.
Furthermore, the transition between the two phases becomes sharper as $\Omega$
increases and/or 
the number of pairs decreases. In passing from the one phase
to the other, we of course pass through the SU(4) phase. 
Note, however, that this
is done smoothly. 
There is no plateau at the SU(4) phase, suggesting no special
stability associated with the maximally alpha-correlated state.

When a particle-hole force is turned on [see Fig. {\ref{fig3}d], some
interesting changes show up. In particular, the pure isoscalar and isovector
phases are never realized. And, furthermore, the transition through the SU(4)
phase is less abrupt. This latter remark may have relevance to real nuclear
systems, which are most likely near to -- but not precisely at -- the SU(4)
limit. Our results suggest that even away from the SU(4) limit, there may be
significant alpha-particle transfer strength, as long as there is a 
sufficiently strong particle-hole force present.

Finally, we close this section by reiterating the
key conclusion of
our analysis of the SO(8) model.  Namely, in the SO(8) model -- a prototypical
model that involves np pairing correlations on the same footing as nn and pp
pairing correlations -- the ground-state wave functions exhibit essential
four-particle correlations.
The fermion SO(8) ground state is constructed in such a way 
that the maximal possible number of fermions form correlated four-particle
$S=0$ $T=0$ structures and the rest form like-particle pairs and/or
an np pair and provide the isospin and spin of the system.

\section{Discussion}
\label{discus}

We have studied neutron-proton pairing correlations within the framework
of two simple, solvable, and somewhat  
realistic nuclear models. In our analysis,
we made extensive use of boson-mapping techniques.
Since the bosons introduced by the (first) mapping represent correlated fermion
pairs, such an approach provides a quite convenient framework in which to
investigate pairing features in fermion systems.

With that in mind, we first studied the isovector-pairing SO(5) model, both
within the framework of boson mean-field methods and also through the use
of genuine fermion-pair approximations.
The latter represent generalizations of
the standard procedures to treat the pairing between like nucleons. We
find that these approaches fail to describe details of neutron-proton
pairing. In particular, the BCS solution does not allow for the coexistence of
like-particle and neutron-proton pairs. As a result, two-nucleon transfer
strengths are not given correctly, even though the approximate energies are
quite close to the exact values. Tiny details of neutron-proton pairing
correlations,  as reflected in double binding energy differences,
are also not well reproduced. Number and $T_z$ projection, by themselves, are
unable to improve the situation.

A nice feature of the boson mapping procedure is that it provides detailed
information on the boson-mapped Hamiltonian, information that can yield useful
clues as to how to introduce the necessary additional correlations. The boson
Hamiltonian that emerged from a mapping of the SO(5) fermion Hamiltonian
contained an attractive pairing force between bosons, suggesting that
correlations between pairs of bosons could be the key missing ingredient to an
improved description of the system. After first exploring the possible use of
boson BCS approximation to accomodate these additional correlations, we
then turned to an alternative treatment based on a second boson
mapping. A simple description of the neutron-proton pairing
problem was achieved in terms of the second bosons, confirming the importance of
boson-boson (or equivalently pair-pair) correlations. Inverting the method,
we were also able to obtain in closed form
the fermion wave functions of the SO(5) model expressed in terms of the
fundamental fermion-pair operators. The wave functions that emerged
clearly showed four-nucleon correlations.

We then carried out a similar study of a somewhat richer model involving
neutron-proton pair correlations based on the algebra SO(8). We focussed our
analysis on the dynamical symmetry limits of this model, where the group
structure could be used to obtain simple closed expressions. Here too we found
that a description limited to fermion-pair (or boson mean-field) correlations
was insufficient and that four-particle correlations were needed to
accurately reproduce the exact results.

Thus, from these model calculations, we conclude that correlations
involving pairs
of fermion pairs, or alternatively quartets of fermions, are
important in the regime of neutron-proton pairing. It is not sufficient only to
pair nucleons. Whenever possible, two nucleon pairs will couple together to form
a $T=0$ $S=0$ (alpha-particle-like) structure.

Qualitatively, this conclusion can be understood as follows. The smallest
``cluster" that can simultaneously accomodate two-neutron pairing correlations,
two-proton pairing correlations and neutron-proton pairing correlations is one
that involves four nucleons -- two neutrons and two protons.  Of course, when
there is an excess of particles of a given type, not all particles can form these
maximally-correlated alpha-like structures. Instead, they remain in like-particle
and/or neutron-proton pairs, appended to the alpha-like {\em condensate}.

The above conclusion is not limited, however, to cases in which all pairing modes
contribute significantly. In the SO$^{\rm S}$(5) limit of the SO(8) model, for
example, which only involves isoscalar np pairing, the ground-state wave function
involves a condensate of $P^{\dagger} \cdot P^{\dagger}$ pairs and thus contains
four-particle correlations. There, however, the fact that the ground state
involves such a quartet structure is a direct reflection of angular momentum
restoration. 

It is important to note, however, that the four-particle
correlated structures that emerge are not exact alpha particles. Nor are they
necessarily the most alpha-like structures available within the model. In the
SO(8) model, for example, the ground state is dominated by the maximal alpha
structure in the SU(4) dynamical symmetry limit only.

A challenging problem that still remains is: What is a good way to incorporate
such four-fermion correlations into nuclear many-body approximation schemes? As
noted above, boson BCS following a boson mapping does not seem to be an
acceptable procedure for incorporating four-nucleon correlations in such models;
fluctuations in the particle number are simply too large. Another possible
approach would be to include the additional correlations through projection.
Isospin projection, in particular, would seem to be important. In the simple
SO(5) model, for example, it leads to the exact solution. It still remains,
however, to develop an appropriate isospin projection technique and to apply it
to realistic nuclear structure problems.

We have put forth two ideas that could perhaps provide the basis for an 
improved
theory with four-particle correlations. One possibility would be to start from
a trial ground-state-wave function in the form of
Eqs. (\ref{gsform})-(\ref{xge0}). The
$S^{\dagger}$ and $P^{\dagger}$ are now collective pair creation operators,
whose structure ideally should be determined variationally (or, less ideally,
from an analysis of simple two-body systems). Such an approach is a
generalization of the generalized-seniority scheme for like nucleons, which is
known to be connected to number-projected BCS theory.

A second possibility worth further investigation is the use of iterative boson
mappings. We have seen that two boson mappings, coupled with a Hartree-Bose
treatment in the second boson space, is a way to build 
a number-projected theory
involving four-nucleon correlations. In the current studies, where spurious
states could be readily identified, such an approach proved extremely useful.
Whether it will continue to prove useful in more realistic applications,
however, where spurious states cannot be so readily separated, still
remains to be investigated.

Other truncation schemes in the shell model should also be studied. Our analysis
of the SO(8) model suggests that the usual truncation schemes built in
terms of separate dynamical symmetry limits for neutrons and
protons (the SU(2) seniority  limit, for
example)  should not be of much use in the general neutron-proton pairing
problem.  A truncation in the SO(5) or SO(8) seniority quantum numbers,
which would then include the necessary isospin correlations, could perhaps be
useful.

\acknowledgements{This work has been supported by the Grant Agency of the
Czech Republic under grant 202/96/1562 and by the U.S. National Science
Foundation under grant \# PHY-9600445. Fruitful discussions with Jacek
Dobaczewski, Jorge Dukelsky and Piet Van Isacker on many aspects of this study
are gratefully acknowledged.}

\end{document}